\documentclass[runningheads]{llncs}

\usepackage[utf8]{inputenc}

\usepackage{graphicx}
\usepackage{cite}
\usepackage{xcolor}
\usepackage{algpseudocode}
\usepackage{algorithm}

\usepackage{amsfonts, amsmath, amssymb}
\usepackage{indentfirst}
\usepackage[all]{xy}
\usepackage{hyperref}

\usepackage{graphicx}
\usepackage{cite}
\usepackage{xcolor}

\usepackage{amsfonts, amsmath, amssymb}
\usepackage{indentfirst}
\usepackage[all]{xy}

\algblock{Input}{EndInput}
\algnotext{EndInput}
\algblock{Output}{EndOutput}
\algnotext{EndOutput}
\newcommand{\Desc}[2]{\State \makebox[2em][l]{#1}#2}

\begin{document}

\title{Mapper Side Geometric Shaping for QAM Constellations in 5G MIMO Wireless Channel with Realistic LDPC Codes}

\titlerunning{Mapper Side Geometric Shaping for QAM Constellations}

\author{
Daniil~Yudakov$^{a,b}$,
Dmitrii~Kolosov$^{a}$,
Evgeny~Bobrov$^{a,b}$
\institute{$^a$ M.V. Lomonosov Moscow State University\\
$^b$ Huawei Technologies, Russian Research Institute, Moscow Research Center}
\thanks{Emails: d.yudakov43@gmail.com, kolosov.dmt@gmail.com, eugenbobrov@ya.ru }\\
}

\authorrunning{Daniil~Yudakov, Dmitrii~Kolosov, Evgeny~Bobrov}

\maketitle

\begin{abstract}
In wireless communication systems, there are many stages for signal transmission. Among them, mapping and demapping convert a sequence of bits into a sequence of complex numbers and vice versa. This operation is performed by a system of constellations~--- by a set of labeled points on the complex plane. Usually, the geometry of the constellation is fixed, and constellation points are uniformly spaced, e.g., the same quadrature amplitude modulation (QAM) is used in a wide range of signal-to-noise ratio (SNR). By eliminating the uniformity of constellations, it is possible to achieve greater values of capacity. Due to the current standard restrictions, it is difficult to change the constellation both on the mapper or demapper side. In this case, one can optimize the constellation only on the mapper or the demapper side using original methodology. By the numerical calculating of capacity, we show that the optimal geometric constellation depends on SNR. Optimization is carried out by maximizing mutual information (MI). The MI function describes the amount of information being transmitted through the channel with the optimal encoding. To prove the effectiveness of this approach we provide numerical experiments in the modern physical level Sionna simulator using the realistic LDPC codes and the MIMO 5G OFDM channels.
\end{abstract}


\section{Introduction}

The performance of communication systems is greatly influenced by the choice of constellation. Both the coordinates and the probabilities of points can be optimized when designing constellations. These approaches are known as geometric and probabilistic shaping, respectively. The geometry of the constellation is usually fixed, e.g. quadrature amplitude modulation (QAM) is used. Probabilistic shaping can still improve the achievable information rate in such cases. An approach of how autoencoders can be used for probabilistic constellation shaping is shown in~\cite{stark2019joint}.   

In optical and wireless communications, several shaping schemes have been proposed recently to match the capacity achieved by the input distribution, thereby increasing the shaping gain. In~\cite{hassan2018applying} probabilistic and geometric shaping schemes for wireless backhaul channels based on their frame error rate performance using soft and hard decision decoding with WiMAX and DVB-S2 LDPC codes are evaluated. Both probabilistic and geometric shaping techniques show significant gains over uniformly distributed symbol transmission for soft decision decoding. However, probabilistic shaping is more challenging because it requires optimizing discrete distributions.

In this paper, we will concentrate on the geometric shaping. This approach is well-studied. For example, geometrically shaped 256-ary constellation in~\cite{chen2018increasing} achieves SNR gains up to 1.18 dB compared to the standard QAM constellations~\cite{i2008bit}. In~\cite{mirani2020low} lattice-based geometrically shaped modulation formats in multidimensional Euclidean space are proposed and also fast and low complexity modulation and demodulation algorithms are described. Recent machine learning techniques have been proposed for geometric shaping~\cite{o2017introduction}. In~\cite{jones2019end} an autoencoder to obtain geometrically shaped constellation is used. 



All these papers suppose that we can change both mapper and demmaper on the transmitter and receiver side. This cannot be realised in existing standards for wireless communication systems. Signal transmission in wireless networks is carried out from the base station to the user equipment (UE). Most UEs work according to the standards already set, for example, 3GPP TS 38.211 V15.4.0~\cite{channelsmodulation}. Changing both of the mapping (uplink) and demapping (downlink) for UEs is difficult, because we need to change the standards themselves.

In this paper, we consider the case when the base station transmit a modified signal, where a sequence of bits is converted into a sequence of complex numbers using non-uniform constellation (NUC) and the user receives this signal and converts it to a sequence of LLRs using standard QAM constellation. This process can be implemented using present technology and does not require any 3GPP standard change. We consider the realistic Low-Density Parity-Check (LDPC) codes and the 5G MIMO OFDM system, providing numerical experiments in the physical communication system level Sionna simulator~\cite{hoydis2022sionna}.

\section{System model}

\begin{figure}
    \centering
    \includegraphics[width=\linewidth]{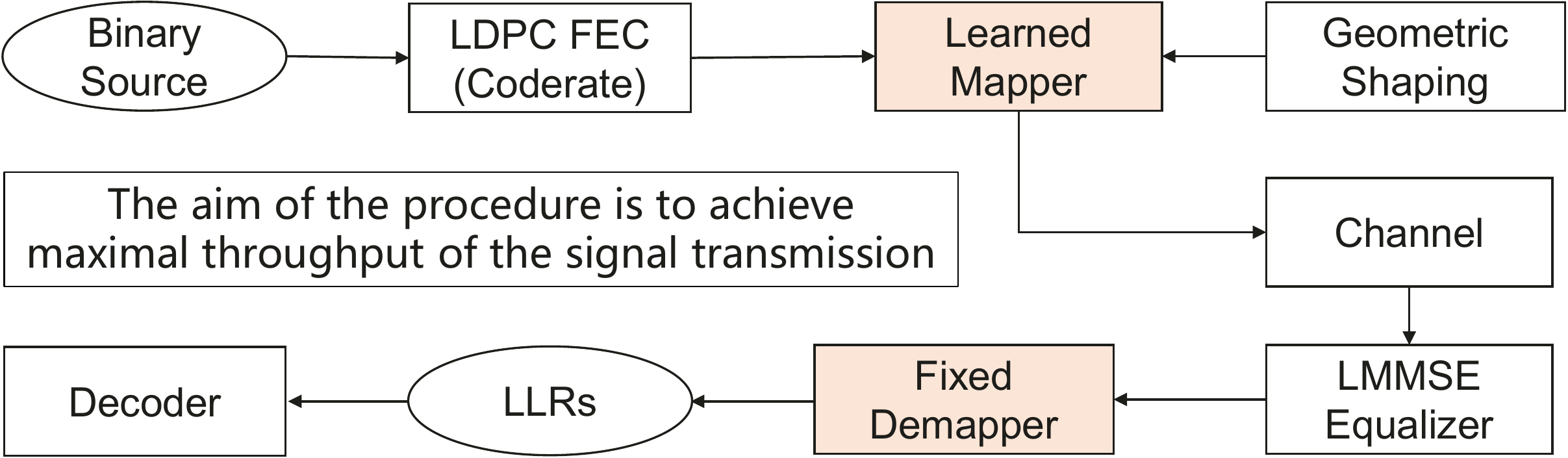}
    \caption{Data transmission workflow with learned mapper and fixed demapper.}
    \label{fig:DTS}
\end{figure}

We consider a realistic point-to-point transmission between a UE and a base station. The system we will configure is shown in the Fig.~\ref{fig:DTS}. The main difference from the usual BICM is the presence of geometric shaping, which changes the mapping procedure.

On the base station side, in the considered channel model, uniformly distributed parity check bits are added to the sequence of bits (binary source) using the LDPC encoder. The bit sequence is then converted into a sequence of complex values using a Mapper. This sequence pass through the AWGN channel.

On the UE side, the LMMSE equalizer reduces the inter-symbol (IS) interference from the received signal (complex values). The demapper converts complex values to LLRs and the decoder finally converts LLRs to bit sequence. The quality of the received signal can be described by the mutual information function.

Our aim is to optimize the modulating part (i.e. the mapper) of this scheme, so that the channel model is as follows
\begin{equation*}
x\xrightarrow{Mapper} s\xrightarrow{Channel} Hs+n\xrightarrow{Equalizer} G(Hs+n)= \tilde{s}\xrightarrow{Demapper} y\xrightarrow{Decoder} \tilde{x} 
\end{equation*}
where $x$ and $\tilde{x}$~--- transmitted and received bit sequences, $s$ and $\tilde{s}$~--- transmitted and received complex signal values, $H$, $G$~--- predefined Channel and LMMSE Equalizer complex matrices; usually $GH\sim I$ (signal does not change with zero noise), $n$~--- complex random noise (e.g. $\mathcal{N}(0,\sigma^2)$), $y$ is a sequence of real values obtained by the Log Likelihood Ratio (LLR) function:
\begin{equation*}
    y_i = llr_i (\tilde{s}) = \log\frac{P(b_i (x)=0|\tilde{s})}{P(b_i(x)=1|\tilde{s})} = \log\left(\frac{\sum_{c\in\mathcal{C}_{i,0}}\exp\left(-\frac{1}{N_0}|\tilde{s}-c|^2\right)}{\sum_{c\in\mathcal{C}_{i,1}}\exp\left(-\frac{1}{N_0}|\tilde{s}-c|^2\right)}\right)
\end{equation*}
where $\mathcal{C}_{i,0}$ and $\mathcal{C}_{i,1}$ are the non-intersecting sets of constellation points. For $c \in \mathcal{C}_{i,0}:$ $i^{th}$ bit equal to 0 and for $c \in \mathcal{C}_{i,1}:$ $i^{th}$ bit equal to 1.

Knowledge of $y = llr(\tilde{s})$ is equivalent to knowledge of $t= P(b(x)=0 | \tilde{s})$ and $1-t=P(b(x)=1 | \tilde{s})$ since $t=\frac{2^y}{2^y+1}$.

Note that in case of fixed demapper LLR function depends on QAM receiver constellation. We have $y_i = llr_i (\tilde{s}, QAM)$. The LLRs can be tabulated once for any new constellation.

\subsection{Mutual Information as a Loss Function}
For an arbitrary channel $X\to Y$ and the input distribution $P_X(x)$ we consider the ensemble $(X,Y)$ with the distribution $P_{(X,Y)}(x,y)$. The correct information measure in this setting is Mutual Information (MI) $I(X;Y)$:
\begin{equation}\label{eq:mi_def}
    I(X;Y):=H(X)-H(X|Y)= \mathbb{E}_{(X,Y)} \left(\log\frac{P_{(X,Y)}(x,y)}{P_X(x) P_Y(y)}\right),
\end{equation}
where $H(X)$ is an Entropy and $H(X|Y)$ is a Conditional Entropy.

The expectation can be computed as a sum or an integral, depending on whether the distribution is discrete or continuous. According to Shannon's theorem~\cite{shannon1948mathematical}, $I(X;Y)$ is a strict upper bound on channel throughput using optimum codes. Modern LDPC codes from the standard are sub-optimal, with a predictable deficit.

The Theorem proved by~\cite{caire1997bit} shows that~\eqref{eq:mi_def} can be calculated as:
\begin{equation}\label{eq:mi_sum}
I(X^{bit} ;Y^{LLR} ) = \sum_{j=1}^m I(B_j; Y) = \sum_{j=1}^m \left( H(B_j)-H(B_j|Y) \right),
\end{equation}
where $m$ is the number of bits transmitted by a given constellation, $B_j$ is a discrete random variable equal to the $j$-th bit, $Y$ is a continuous random variable equal to a received point on a complex plane. As long as the transmitter and receiver use the same constellation, this formula is correct for any constellation and any type of noise. This means that we can use an error-correcting code for the whole set of bits without losing any information.


\subsection{Cross-entropy for Neural Networks}

To calculate $I(X;Y)$~\eqref{eq:mi_sum} we need to calculate each term of sum: 
\begin{equation}\label{eq:mi_term}
    I(B_j;Y)= H(B_j)-H(B_j|Y)
\end{equation}

The first term in~\eqref{eq:mi_term} can be easily computed due to the fact of equal probability of zero and one in bit sequence:
\begin{equation*}
    H(B_j )=- \frac{1}{2}\log\frac{1}{2}-\frac{1}{2}\log\frac{1}{2} = 1
\end{equation*}

Now we need to calculate the second term in~\eqref{eq:mi_term}: $H(B_j|Y)$. For that, let the variable $t = t(\tilde{s})$ be a probability that the transmitted bit is zero, $b_j(x)=0$, with a condition of received complex value $\tilde{s}$: $t = t(\tilde{s}) = P(b_j (x)=0 | \tilde{s})$. Then the probability that the one is transmitted $P(b_j (x)=1 | \tilde{s})=1-t$ and
\begin{multline*}
    H(B_j|Y) 
    = -\int P(y,b_j (x)=0) \log P(b_j (x)=0|y)  dy - \\ - \int P(y,b_j (x)=1) \log P(b_j (x)=1 | y)  dy 
    = \\ = -\int P(y)P(b_j (x)=0 | y) \log P(b_j (x)=0 | y)  dy - \\ - \int P(y)P(b_j (x)=0 | y) \log P(b_j (x)=0 | y)  dy  
    = \\ = -\int P(y)  (t \log t+(1-t) \log (1-t) dy)
    = \int P_Y (y) H_2 (t) dy,
\end{multline*}
where $H_2(t) = t\log\frac{1}{t}+(1-t)\log\frac{1}{1-t}$, is known as \textit{binary cross-entropy}.

The function $H(B_j|Y)$ is used in the current optimization procedure.




\subsection{Mapper Side Constellation Optimization Features}

Our approach involves only constellation changing on the mapper side. This entails a change to the MI function. Mapper constellation is responsible for distribution $P_Y(y)$. On the other hand, fixed demmaper QAM constellation is responsible for function $H_2(t)$. 

\begin{algorithm}

\caption{An Adam algorithm for mapper constellation optimization}\label{alg:cap}
\begin{algorithmic}
  \Input
  \Desc{\textit{SNR}, $Coderate$, $QAM\_Constallation$}
  \EndInput
  \Output
  \Desc{$Mapper\_Constellation$}
  \EndOutput

\State $Mapper\_Constellation \gets QAM\_Constallation$
\State $Demapper\_Constellation \gets QAM\_Constallation$

\For{$t = 1:T$} \Comment{$T$ is the number of Adam epochs}
        \State $bits \gets BinarySource()$ \Comment{Random sequence generation}
        \State $s \gets Mapper(bits, Mapper\_Constellation)$ 
        \State $\tilde{s} \gets AWGN\_Channel(s, SNR)$ 
        \State $llrs \gets Demapper(\tilde{s}, Demapper\_Constellation)$ 
        \State $loss \gets BinaryCrossentropy(bits, llrs)$ 
        \State $Mapper\_Constellation \gets AdamUpdate(Mapper\_Constellation, loss)$
\EndFor\\
\Return{$Mapper\_Constellation$}
\end{algorithmic}
\end{algorithm}

\begin{figure}
    \centering
    \includegraphics[width=\linewidth]{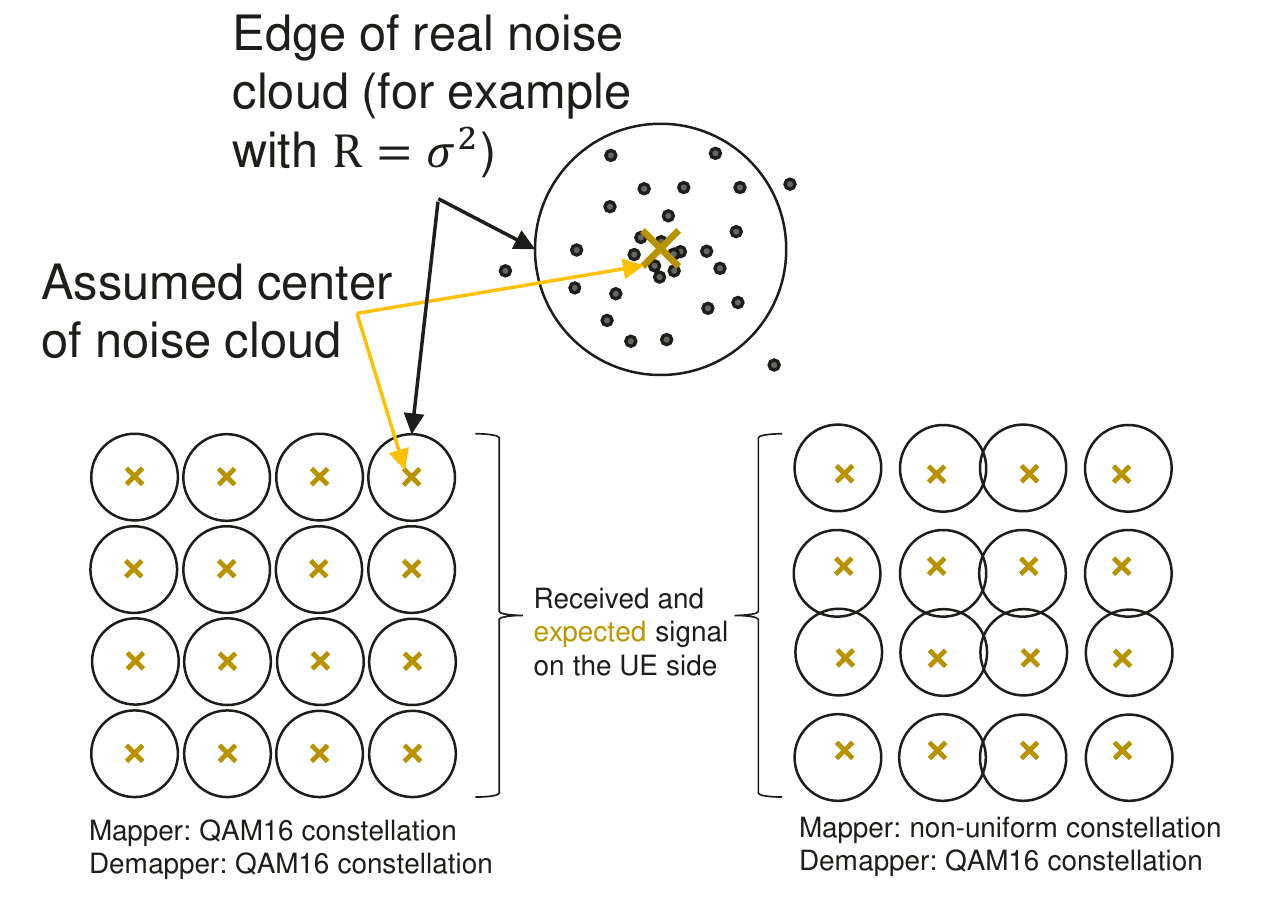}
    \caption{Illustration of user-side changes. Here UE considers that received signal distributed as Gaussian noise around QAM constellation points (around yellow cross). Demapping procedure calculate LLR's based on this assumption. But the real signal has different distribution for different mapper constellation. On the pictures these distributions represented as the circles. We can say that the mathematical expectation of each group of received points connected to a fixed mapper constellation point will be in the center of some circle. Radius $R$ of this circle shows dispersion of this noise. Due to the Gaussian nature of noise, we can't construct a strict edge of noise (some points may be outside the circle).}
    \label{fig:NUC}
\end{figure}

We can explain these facts as follows: $y$ is the signal received by the user after passing through the channel. The  distribution $P_Y(y)$ describes the probability of detecting a given signal in a certain area. For the AWGN channel, this distribution is represented as noise clouds around the constellation points. Thus, if we move the mapper constellation points, the distribution $P_Y(y)$ itself changes (See Fig.~\ref{fig:NUC}). The movement of one circle leads to an increase in interference with some circles and a decrease with others. The MI function helps us move it properly. On the other hand, demapper constellation points show how this signal $y$ is converted to LLR's. As described above, the $t(y)$ function is responsible for this. It doesn't change because the demapper constellation doesn't change. So, to minimize the MI function, we need to maximize $\int P_Y (y) H_2(t(y)) dy$ with fixed function $t(y)$.

\section{Simulation Results}

\begin{figure}
    \includegraphics[width=0.5\textwidth]{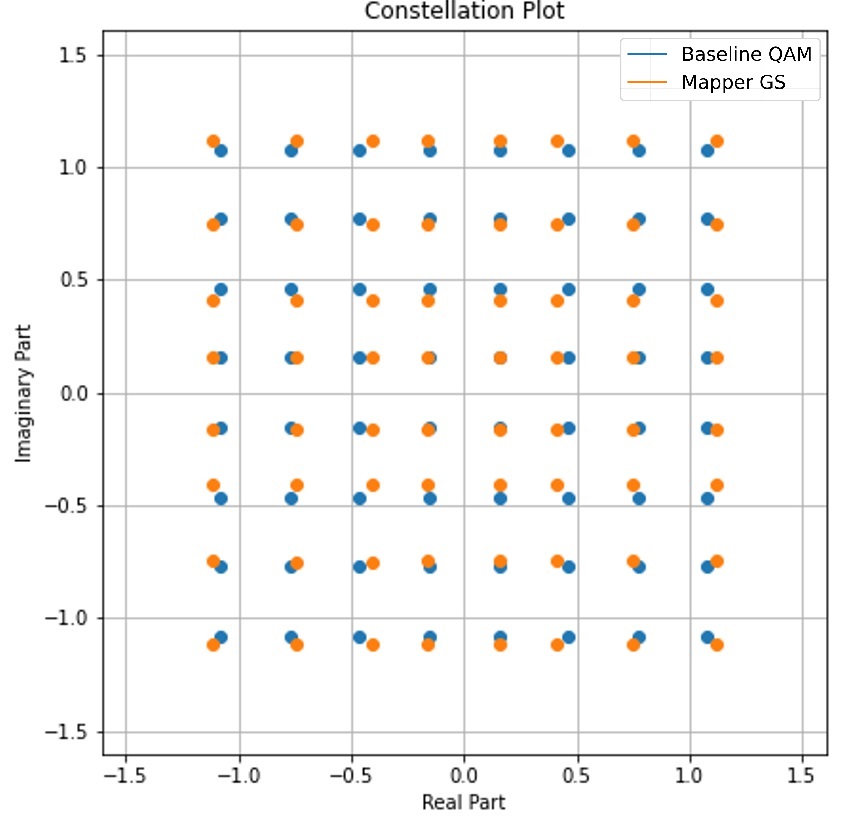}\;
    \includegraphics[width=0.5\textwidth]{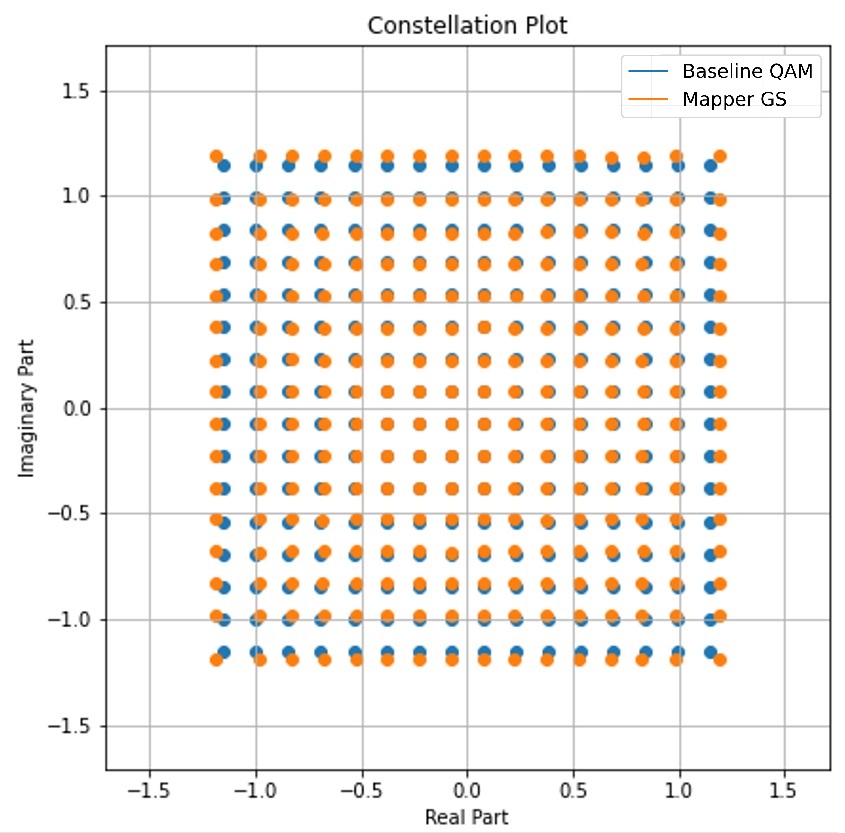}
    \caption{Examples of shaped constellation. The blue points represent the base QAM constellation. The orange points represent the shaped constellation.}
    \label{fig:Constellations}
\end{figure}

\begin{figure}
    \centering
    \includegraphics[width=\linewidth]{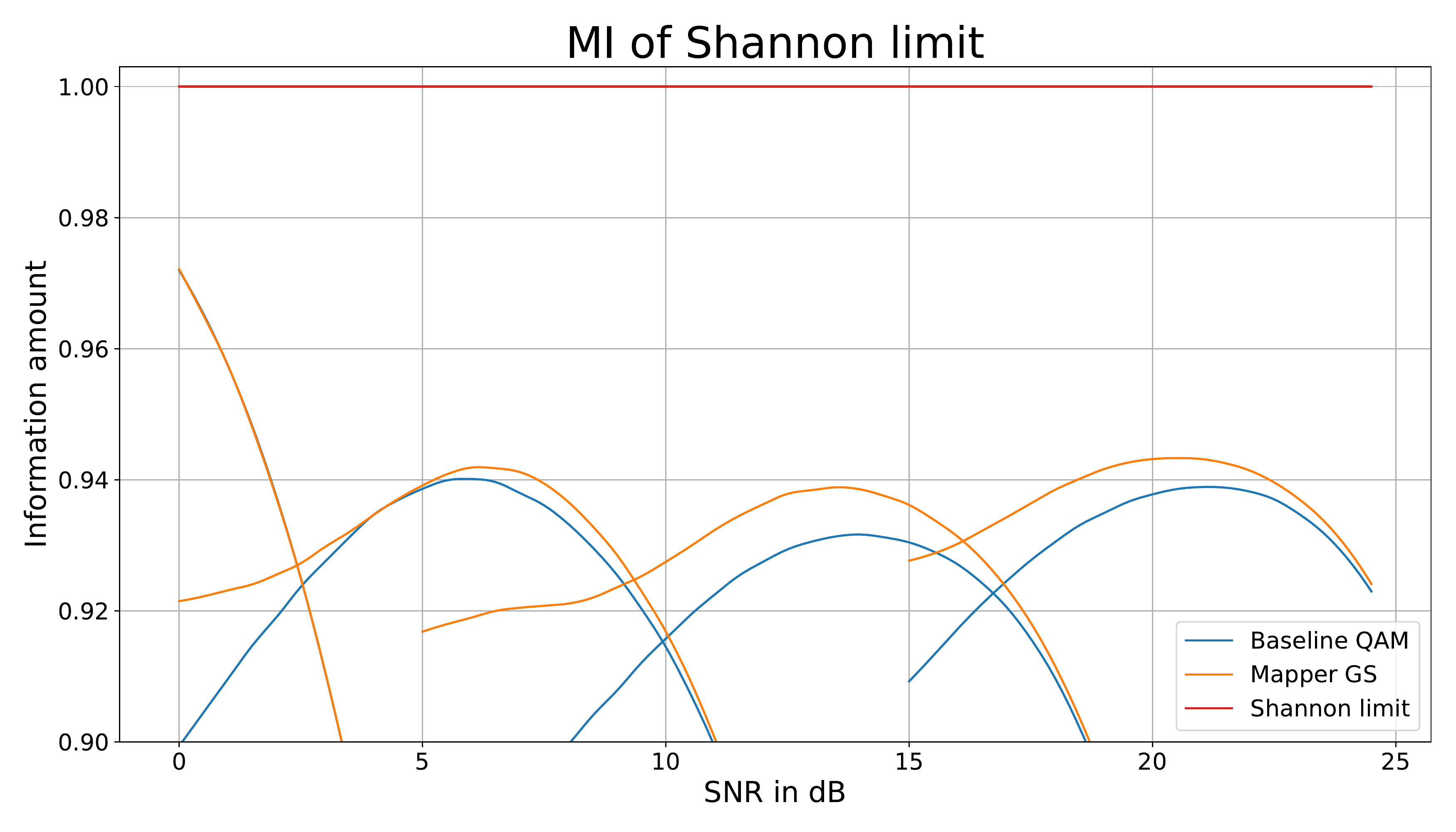}
    \caption{ Mutual Information. The $x$-axis represents SNR in dB. The $y$-axis represents the percentage of additional information we can transmit to the user, compared to the Shannon limit. Here the blue line shows percentage of the base QAM modulation transmission. The orange line shows the percentage of transmission with the shaped constellation.}
    \label{fig:MI}
\end{figure}

The tests were performed on the Sionna~\cite{hoydis2022sionna} simulation platform. The first part of the simulations was conducted for the simplified channel. Here we obtained the optimal constellations by MI maximization (See some examples of constellation on Fig.~\ref{fig:Constellations}). Constellation training was done using the Adam optimizer. For each SNR, the amount of information transmitted in the cases of uniform and non-uniform constellations was calculated (See Fig.~\ref{fig:MI}). Here we see that SAP constellations have up to 1.5\% gain in MI for high SNR (Signal to Noise Ratio) values.

\begin{figure}
    \includegraphics[width=\linewidth]{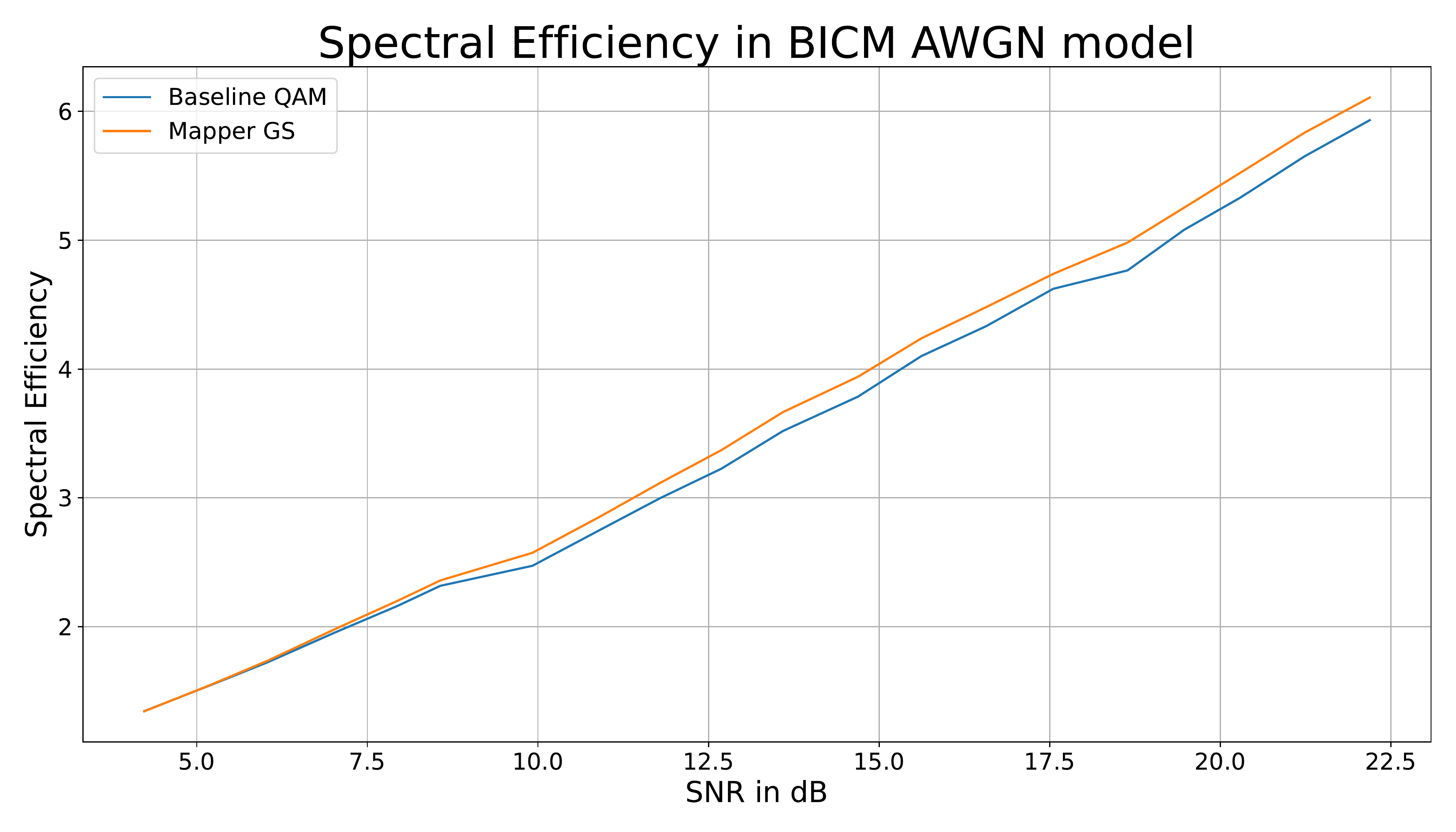}
    \caption{Spectral Efficiency. The $x$-axis represents SNR in dB. The $y$-axis represents the amount of information we can transmit to the user due to the unit period of time. Here the blue line shows amount of information of the base QAM modulation transmission. The orange line shows the amount of information of transmission with the shaped constellation.}
    \label{fig:SE}
\end{figure}

The resulting constellations were tested in more complex scenarios. First, we considered a scenario where constellations were merged with the LDPC encoder. Sionna simulator considers LDPC codes from standard 3GPP TS 38.211 V17.0.0 (2021-12). The result can be seen in Fig.~\ref{fig:SE}. Here “Baseline QAM” is the algorithm with a uniform mapper constellation, “Mapper GS” is the algorithm with a non-uniform mapper constellation that maximizes the MI function.
    
Spectral efficiency (in bits per unit of time) is an indicator of bandwidth efficiency. In our case, it is defined as
\begin{equation*}
    SE = (1-BLER)\cdot Coderate \cdot num\_bits
\end{equation*}
where $BLER$ is the Block Error Rate (probability of error when transmitting a single LDPC block),
$Coderate$ is MCS-based encoding rate,
$num\_bits$ is the number of bits encoding the constellation. For example, for QAM16: $num\_bits=log_2(16)=4$. The value of $num\_bits$ also depends on the MCS.

The values of $Coderate$ and $num\_bits_{qam}$ are obtained from 3GPP standards. The value of $BLER$ was obtained by simulation.

\begin{figure}[H]
    \includegraphics[width=\linewidth]{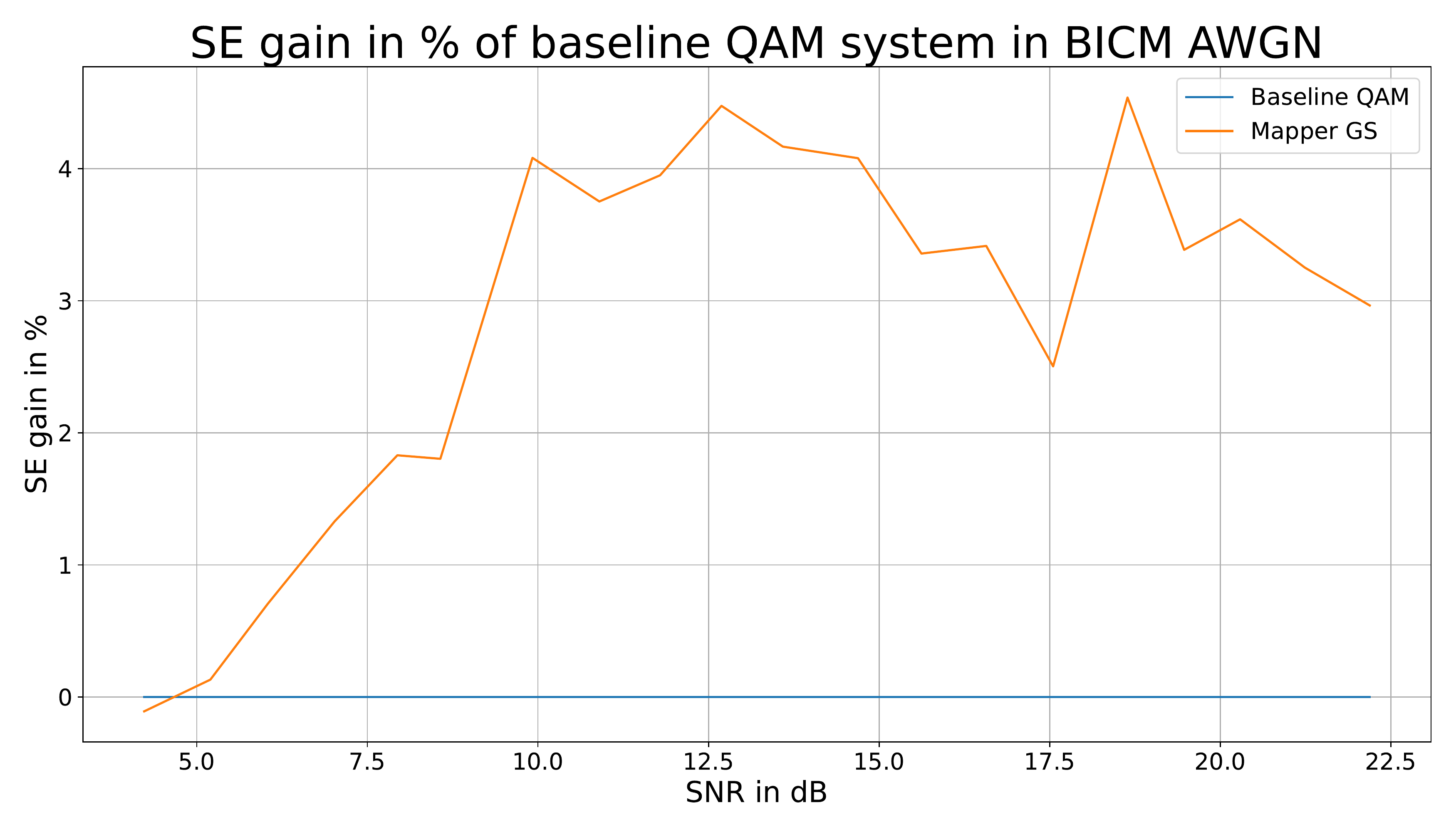}
    \caption{Spectral Efficiency Gain. The $x$-axis represents SNR in dB. The $y$-axis represents the percentage of information we can transmit to the user, compared to the  base QAM modulation transmission. The orange line shows the percentage of transmission with the shaped constellation.}
    \label{fig:SE_gain}
\end{figure}

Compared to the old approach, the graph of the gains 
$$SE_{gain}=\frac{SE_{after}}{SE_{before}}-1$$ 
over the basic algorithm is more descriptive (Fig.~\ref{fig:SE_gain}). 

For this series of experiments we get a gain of up to 4\%.

The second series of experiments was conducted with taking into account Orthogonal Frequency-Division Multiplexing (OFDM) modulation. The architecture of the system consists of Forward Error Correction LDPC Code, Bit Interleaver, Resource Grid Mapper, Least-Squares Channel Estimator, Nearest Neighbor Demapper, LMMSE Equalizer, OFDM Modulator and NUC Constellations. 

\begin{figure}[H]
\includegraphics[width=1\linewidth]{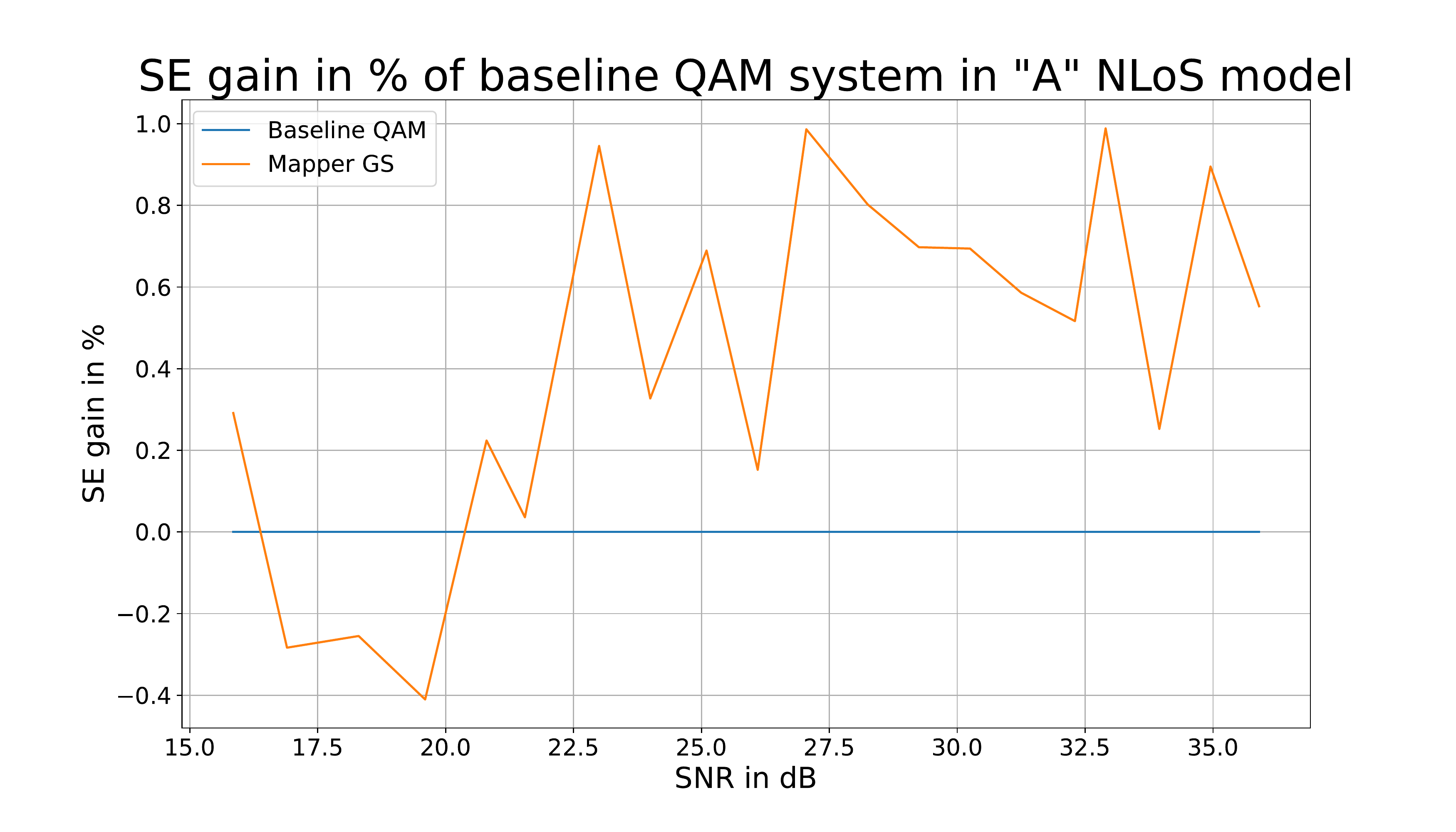}
\caption{Spectral Efficiency gain for 3GPP wireless NLoS A channel model.}
\label{fog:SE_ModelA}
\end{figure}

\begin{figure}[H]
\includegraphics[width=1\linewidth]{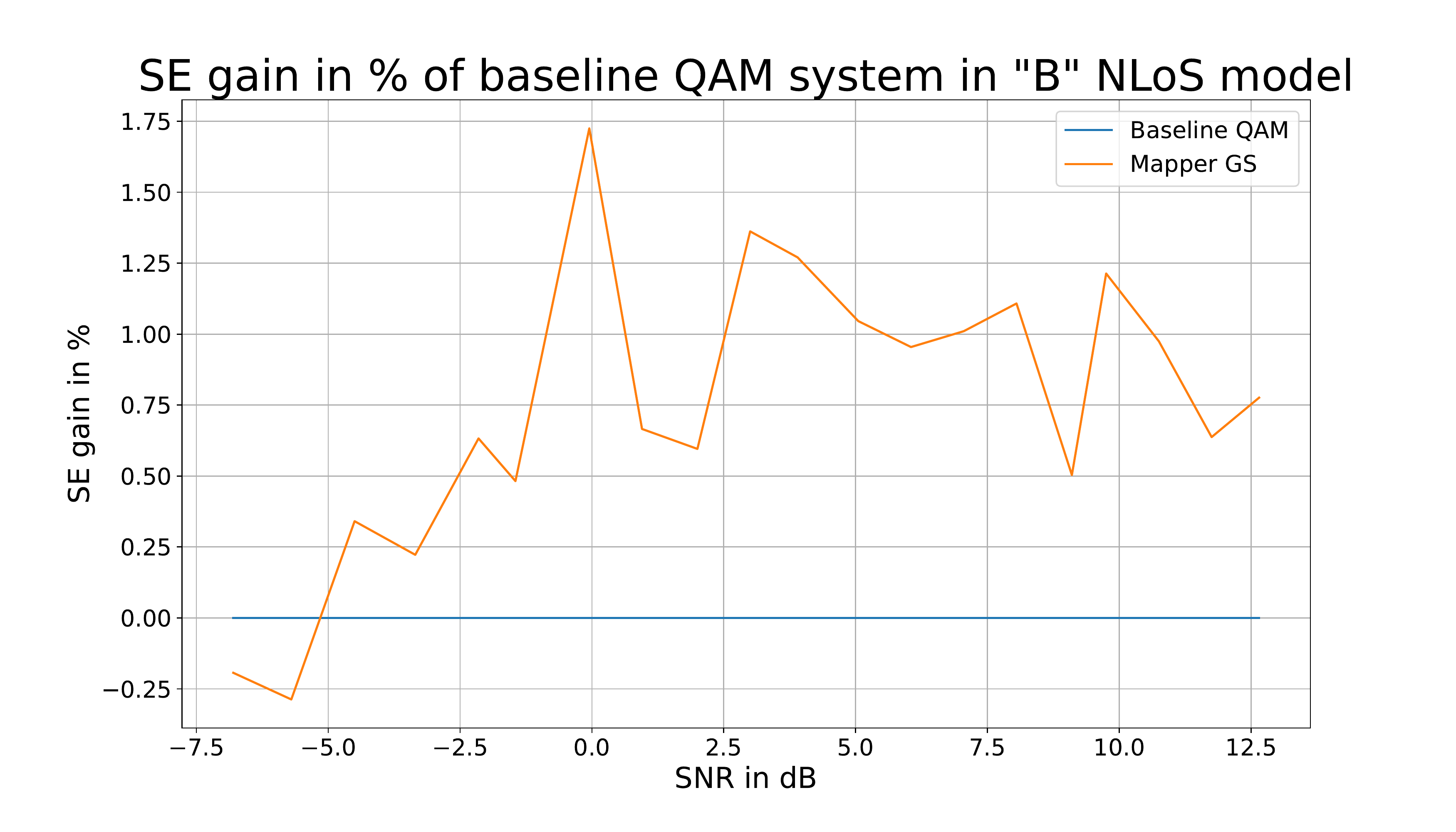}
\caption{Spectral Efficiency gain for 3GPP wireless NLoS B channel model.}
\label{fog:SE_ModelB}
\end{figure}

\begin{figure}[H]
\includegraphics[width=1\linewidth]{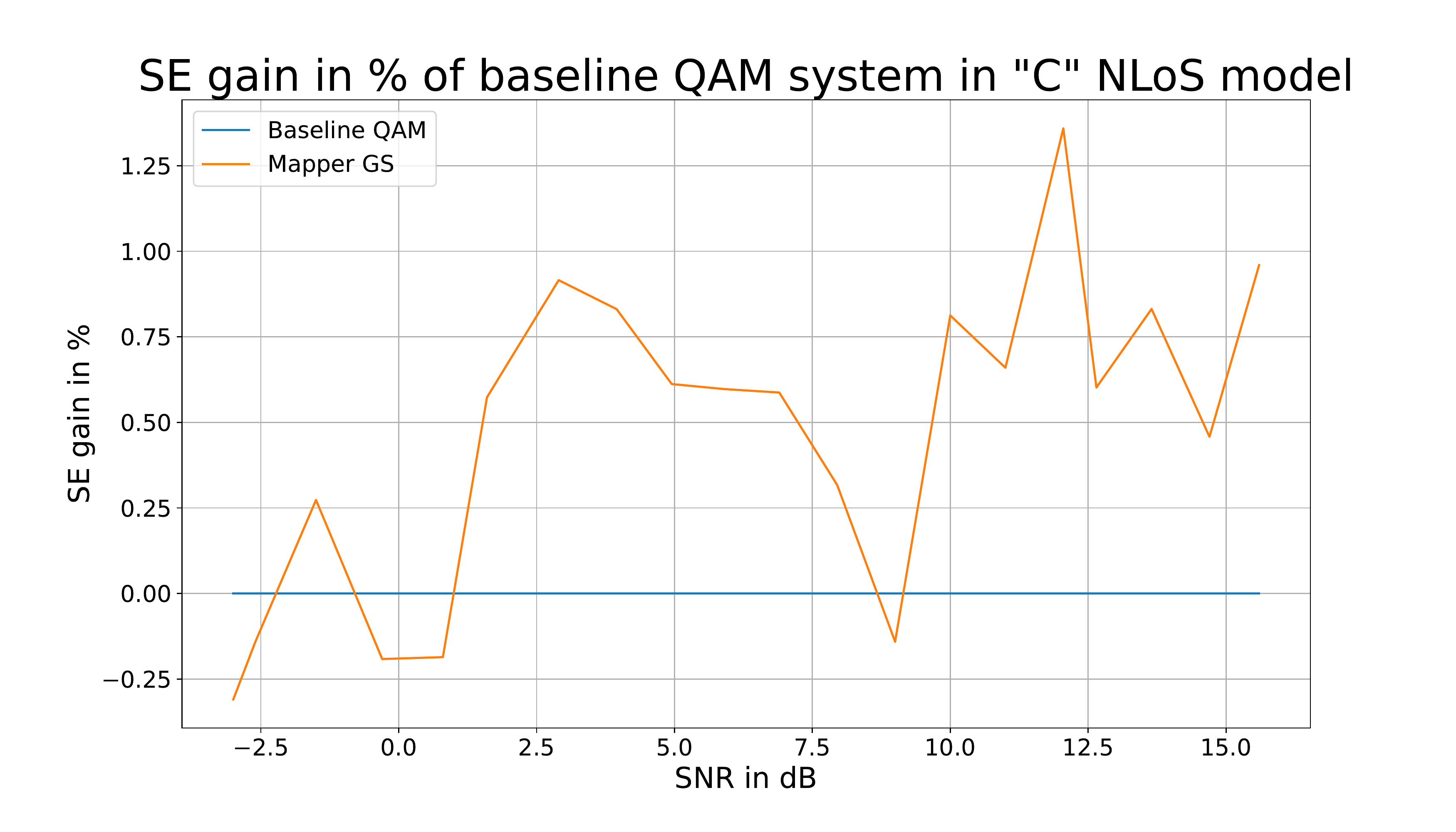}
\caption{Spectral Efficiency gain for 3GPP wireless NLoS C channel model.}
\label{fog:SE_ModelC}
\end{figure}

\begin{figure}[H]
\includegraphics[width=1\linewidth]{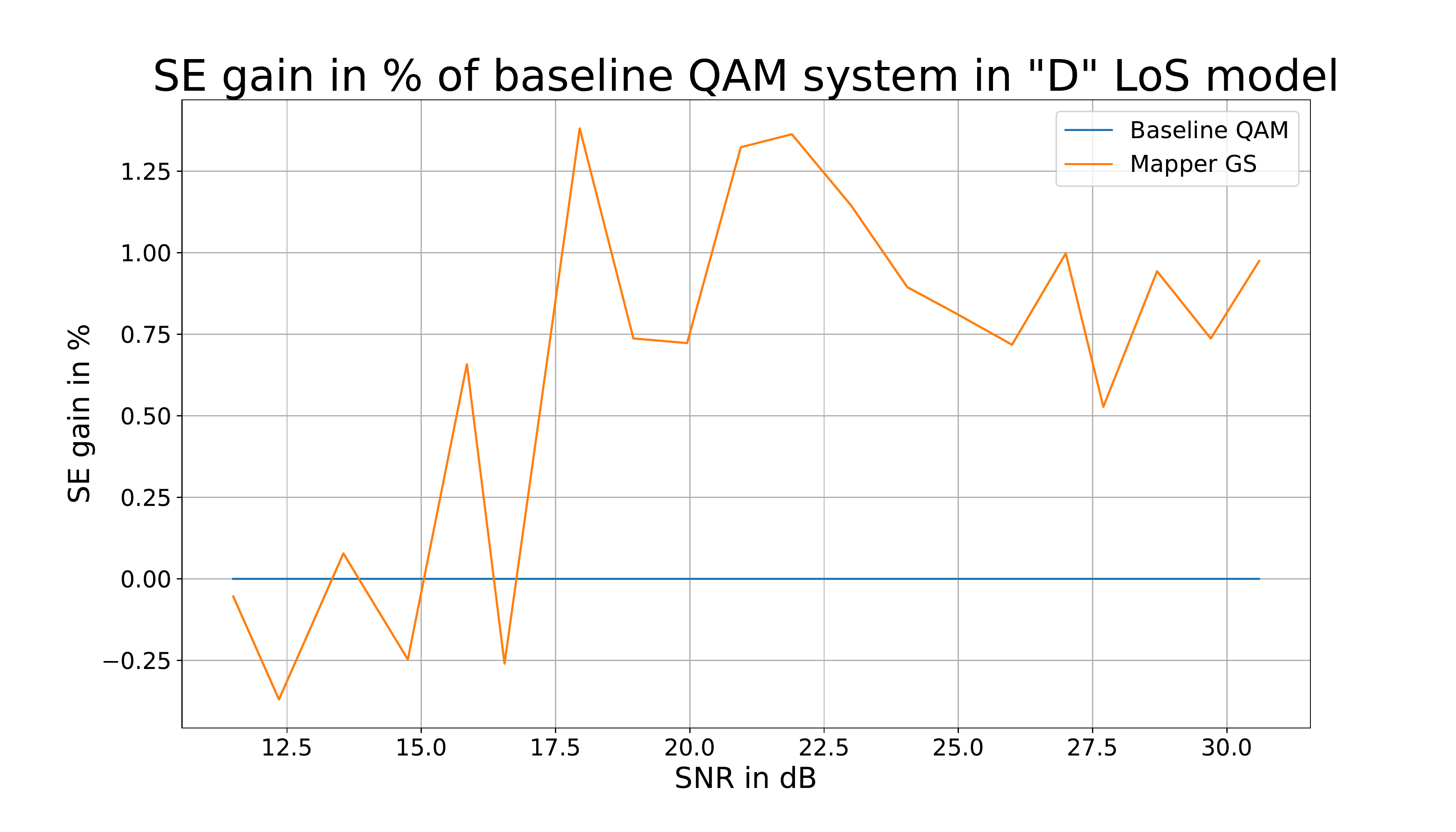}
\caption{Spectral Efficiency gain for 3GPP wireless LoS D channel model.}
\label{fog:SE_ModelD}
\end{figure}

\begin{figure}[H]
\includegraphics[width=1\linewidth]{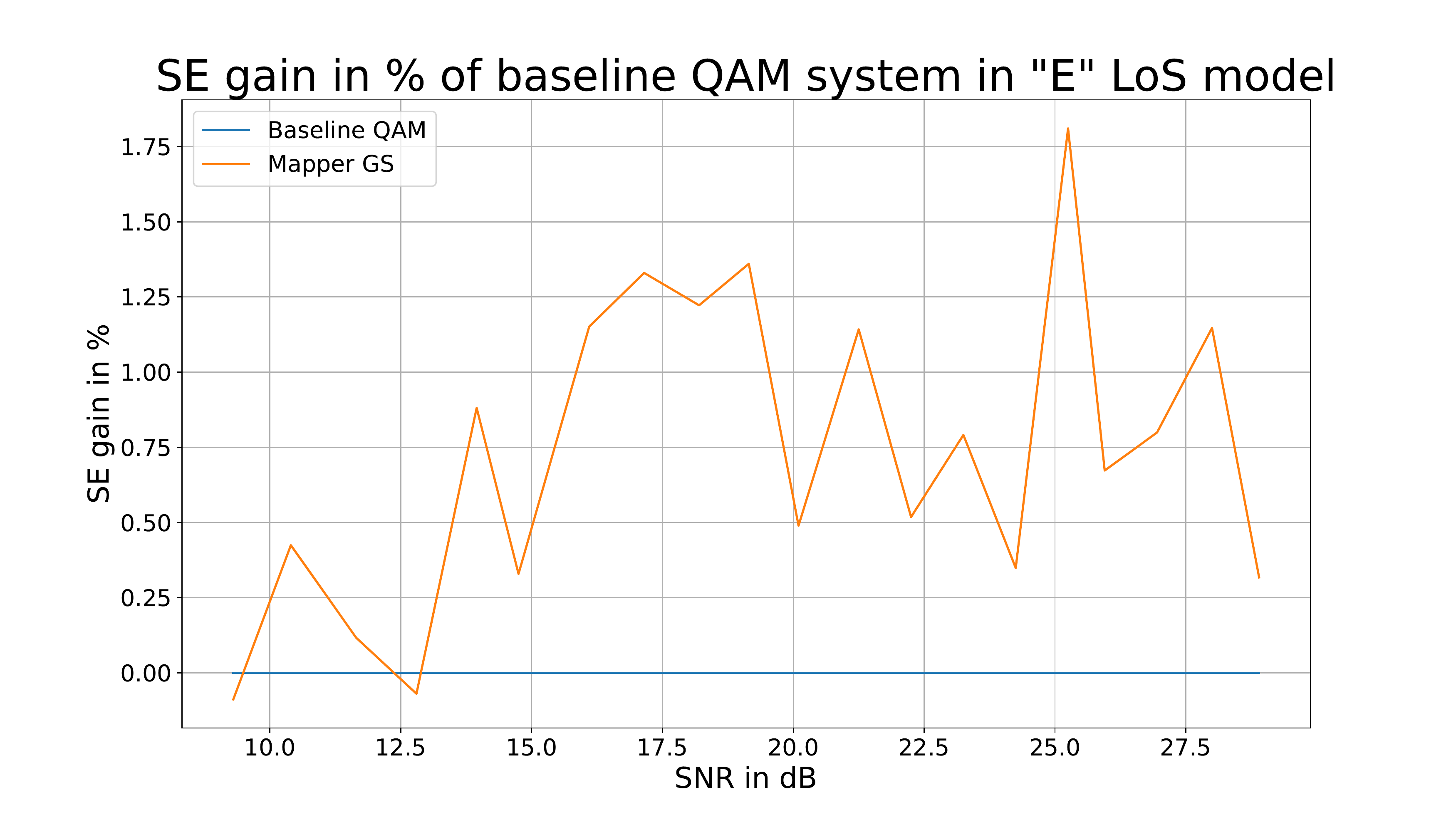}
\caption{Spectral Efficiency gain for 3GPP wireless LoS E channel model.}
\label{fog:SE_ModelE}
\end{figure}

The system uses different 3GPP wireless  Non-Line-of-Sight (NLoS) Clustered Delay Line (CDL) channel models: A, B, C; and Line-of-Sight (LoS) CDL channel models: D, E (Figs.~\ref{fog:SE_ModelA},~\ref{fog:SE_ModelB},~\ref{fog:SE_ModelC},~\ref{fog:SE_ModelD},~\ref{fog:SE_ModelE})~\cite{barb2019influence}. The time models are simulated in real time domain considering inter-symbol (IS) and inter-carrier (IC) interferences, while frequency models are simulated straight away in frequency domain without taking into account IS and IC interferences.

For these series of experiments we get a gain of up to 1.75\%.

\section{Conclusion}

This paper presents the solution for the problem of constellation shaping with the condition that it can be changed only on the transmitter side. Constellations optimized on a simple model improve the spectral efficiency function on more complicated models that take into account different encoders. It has been shown that such an approach can have a gain of about 4\% for simple AWGN channel models with LDPC code and about 1.5\% for OFDM LoS and NLoS models.

\bibliographystyle{splncs04}
\bibliography{mybibliography}

\begin{thebibliography}{10}
\providecommand{\url}[1]{\texttt{#1}}
\providecommand{\urlprefix}{URL }
\providecommand{\doi}[1]{https://doi.org/#1}

\bibitem{barb2019influence}
Barb, G., Otesteanu, M.: On the influence of delay spread in tdl and cdl
  channel models for downlink 5g mimo systems. In: 2019 IEEE 10th Annual
  Ubiquitous Computing, Electronics \& Mobile Communication Conference
  (UEMCON). pp. 0958--0962. IEEE (2019)

\bibitem{caire1997bit}
Caire, G., Taricco, G., Biglieri, E.: Bit-interleaved coded modulation. In:
  Proceedings of ICC'97-International Conference on Communications. vol.~3, pp.
  1463--1467. IEEE (1997)

\bibitem{channelsmodulation}
Channels, N.P.: Modulation (release 15), v15. 4.0, document ts 38.211, 3gpp,
  dec. 2018

\bibitem{chen2018increasing}
Chen, B., Okonkwo, C., Hafermann, H., Alvarado, A.: Increasing achievable
  information rates via geometric shaping. In: 2018 European Conference on
  Optical Communication (ECOC). pp.~1--3. IEEE (2018)

\bibitem{i2008bit}
i~Fabregas, A.G., Martinez, A., Caire, G., et~al.: Bit-interleaved coded
  modulation. Foundations and Trends{\textregistered} in Communications and
  Information Theory  \textbf{5}(1--2),  1--153 (2008)

\bibitem{hassan2018applying}
Hassan, N.U., Xu, W., Kakkavas, A.: Applying coded modulation with
  probabilistic and geometric shaping for wireless backhaul channel. In: 2018
  IEEE 29th Annual International Symposium on Personal, Indoor and Mobile Radio
  Communications (PIMRC). pp.~1--5. IEEE (2018)

\bibitem{hoydis2022sionna}
Hoydis, J., Cammerer, S., Aoudia, F.A., Vem, A., Binder, N., Marcus, G.,
  Keller, A.: Sionna: An open-source library for next-generation physical layer
  research. arXiv preprint arXiv:2203.11854  (2022)

\bibitem{jones2019end}
Jones, R.T., Yankov, M.P., Zibar, D.: End-to-end learning for gmi optimized
  geometric constellation shape. In: 45th European Conference on Optical
  Communication (ECOC 2019). pp.~1--4. IET (2019)

\bibitem{mirani2020low}
Mirani, A., Agrell, E., Karlsson, M.: Low-complexity geometric shaping. Journal
  of Lightwave Technology  \textbf{39}(2),  363--371 (2020)

\bibitem{o2017introduction}
O’shea, T., Hoydis, J.: An introduction to deep learning for the physical
  layer. IEEE Transactions on Cognitive Communications and Networking
  \textbf{3}(4),  563--575 (2017)

\bibitem{shannon1948mathematical}
Shannon, C.E.: A mathematical theory of communication. The Bell system
  technical journal  \textbf{27}(3),  379--423 (1948)

\bibitem{stark2019joint}
Stark, M., Aoudia, F.A., Hoydis, J.: Joint learning of geometric and
  probabilistic constellation shaping. In: 2019 IEEE Globecom Workshops (GC
  Wkshps). pp.~1--6. IEEE (2019)

\end{thebibliography}

\end{document}